\documentstyle[jkas]{article}

\beginpage{93}
\endpage{99}
\year{2012}\volume{45}\month{August}
\runningauthor {S. Youn \& S. Kim}
\runningtitle{POWER SPECTRUM ANALYSIS OF THE OMC1 IMAGE AT 1.1MM WAVELENGTH}
\month{August} \year{2012} \volume{45} \issueno{4}
\beginpage{93}\endpage{99}
\date{Received June 7, 2012; Revised July 29, 2012; Accepted August 6, 2012}

\begin{document}

\title{POWER SPECTRUM ANALYSIS OF THE OMC1 IMAGE AT 1.1MM WAVELENGTH}
\author{Soyoung Youn$^{1}$, and Sungeun Kim$^{1}$}
\address{$^1$ Department of Astronomy and Space Science, Sejong University, Gunja-dong, Gwangjin-gu, Seoul 143-747, Korea\\ {\it E-mail : sek@sejong.ac.kr}}

\address{\normalsize{\it (Received June 7, 2012; Revised July 29, 2012; Accepted August 6, 2012)}}
\offprints{S. Kim}
\abstract{We present a 1.1mm emission map of the OMC1 region observed with AzTEC, a new large-format array composed of 144 silicon-nitride micromesh bolometers, that was in use at the James Clerk Maxwell Telescope (JCMT). These AzTEC observations reveal dozens of cloud cores and a
tail of filaments in a manner that is almost identical to the submillimeter continuum emission of the entire OMC1 region at 450 and 850 $\mu$m.
We perform Fourier analysis of the image with a modified periodogram and the density power spectrum, which provides the distribution of the length scale of the structures, is determined. The expected value of the periodogram converges to the resulting power spectrum in the mean squared sense. The present analysis reveals that the power spectrum steepens at relatively smaller scales. At larger scales, the spectrum flattens and the power law becomes shallower. The power spectra of the 1.1mm emission show clear deviations from a single power law. We find that at least three components of power law might be fitted to the calculated power spectrum of the 1.1mm emission. The slope of the best fit power law, $\gamma \approx -2.7$ is similar to those values found in numerical simulations. The effects of beam size and the noise spectrum on the shape and slope of the power spectrum are also included in the present analysis. The slope of the power law changes significantly at higher spatial frequency as the beam size increases.}
\maketitle

\section{INTRODUCTION}

The Orion Nebula is located in the northern part of the Orion A molecular cloud and consists of OMC1, OMC2, OMC3, and OMC1-S. These components are smoothly connected and form integral shaped filaments seen in continuum observations at sub-millimeter, infrared, and optical wavelengths. The M42, Orion Nebula, is known to be the best studied star-forming region in the nearest giant molecular cloud (GMC) where massive stars have formed and has been extensively studied at all wavelengths. OMC1 in the Orion A molecular cloud possesses the most massive component containing at least 2,000 young stellar objects (YSOs) (O'Dell 2001) and is strongly connected to the surrounding gas. There is also a report on the VLA NH$_3$ observations and revelations of the filamentary and clumpy structure of OMC1 in the Orion A molecular cloud (Wiseman \& Ho 1996, 1998). OMC1 has an on-going massive star formation and strong interactions between gas and young stellar objects (YSOs), their bipolar stellar jets, and outflows (Mann \& Williams 2010; Bally et al. 2011). As we have mentioned above, the complex and filamentary structure of the Orion A molecular cloud, especially the Orion Integral Shaped Filament was revealed by sub-millimeter continuum observations by Lis et al. (1998) and Johnstone \& Bally (1999). The condensation mass spectrum was also explored and temperature distribution was examined (Johnstone \& Bally 1999). Mookerjea et al. (2000) found that the coldest clump in the Orion A molecular cloud had a temperature of about 15 K.
These cold clumps can be traced at the continuum observations using the observations at mm wavelength. Despite various observations of the Orion A molecular cloud over a wide range of wavelengths for over a decade including the SEST SIMBA (Nyman et al. 2001), thermal continuum emission at 1.1mm has been observed for the first time only recently, owing to the development of the Astronomical Thermal Emission Camera (AzTEC). This AzTEC camera is composed of an array of 144 nitride micro-mesh composite bolometers (Wilson et al. 2008). This camera was originally designed for a millimeter-wavelength bolometer camera to be installed on the Large Millimeter Telescope (LMT). But in year 2005 and 2006, it was installed on the James Clerk Maxwell Telescope (JCMT) in Hawaii, Mauna Kea and performed biased and unbiased surveys, of the northern sky.

The present study analyzes the 1.1mm AzTEC observations of the OMC1 components and reports the fact that the 1.1mm emission from OMC1 is distributed in a manner almost similar to that of the 850 $\mu$m emission (Johnstone \& Bally 1999) observed with the Submillimeter Common-User Bolometer Array (SCUBA) on the JCMT. In the present analysis, we performed a power spectrum analysis of the 1.1mm intensity map which is dominated by thermal emission from cold dust in OMC1 in the Orion A molecular filaments. Since the dust grains are coupled with gas, elucidating the structure of the dust emission will also play a crucial role in understanding the dynamics and the structure of the gas. In general, the density fluctuations of the interstellar dust (Draine \& Lazarian 1998) occur in the cold and dense interstellar medium (ISM).

The power spectrum analysis using the Fourier conversion of the given images is especially important because they can reveal the statistical properties of structures which may be present across the images. The generation of the hierarchical structures or fractal structures (Falgarone et al. 1991; Elmegreen \& Falgarone 1996) often presents across a range of spatial scales in the ISM. It has also been suggested that this could be attributed to turbulent emission since this can increase gas density and initiate star forming instabilities (Mac Low 2004; Burkert 2006). However, the large scale properties of turbulence in the ISM, including those of the gas and dust grains are still poorly understood. But there has been an explosion of research in this area. In general, the energy input from star formation is thought to be a major driving force for the turbulence in the ISM (Scalo \& Elmegreen 2004). It is also well known that measurements of the slopes of the power spectra (Crovisier \& Dickey 1983; Gautier et al. 1992) from the intensity fluctuations in the turbulent medium can provide an insight into the nature of scales present in the hierarchical or non-hierarchical structures (Blitz \& Williams 1997; Hartmann 2002) in the interstellar medium (Brunt 2010). Thus, the results from the present studies of the power spectrum analysis can reveal the nature of the structures seen in the 1.1mm image of OMC1 in the Orion A molecular cloud and filaments. The results of this present analysis are presented in Section 3 and 4. Section 2 provides procedures of image processing and the resultant data.

\section{OBSERVATION AND DATA PROCESSING}
\label{s:observe}

In this section, we describe the processing of the 1.1mm AzTEC survey data sets. The observations of OMC1 were performed in raster-scan mode between 28 December 2005 and 1 January 2006 using the AzTEC mounted on the JCMT, giving a beam size of 18 arcsec and a scan velocity of 120 arcsec s$^{-1}$. A total integration time of 102.9 seconds was used. From a total of ten scan maps, five scans were combined, covering an observing area of 655.35 arcmin$^{2}$, which corresponds to about 3.14 $\times$ 3.14 pc$^{2}$. Average RMS was measured to be 0.22 mJy. The observed region was centered on RA=$05^{\rm h}$$35^{\rm m}$$14.4^{\rm s}$, DEC=$-05^{\circ}$$22^{\prime}$$32.3^{\prime\prime}$ in J2000. AzTEC suffered a pointing error of $2^{\prime\prime}$. Beam maps were made from observations of Uranus with a mean flux density of 42.2 Jy at 1.1mm
during the JCMT observing run (Wilson et al. 2008).

A pointing model was generated using CRL618, a post-asymptotic giant branch (post-AGB) star with a mean flux density of 2.7 Jy at 1.1mm. The intensity calibration errors were estimated to be about $\pm$6-13$\%$ (Wilson et al. 2008). The raw bolometer data was calibrated and edited in the IDL AzTEC software (Perera et al. 2008; Scott et al. 2008; Wilson et al. 2008; Austermann et al. 2009). Each data set was regarded as being in the time sector using the data storage method. Since all the data was stored as a function of time, they were also referred to as time-stream data. The AzTEC data set was also in the spatial sector. Each observed data set \textit{i}, $M_i$, was composed of the atmospheric signal, $A_i$, the astronomical signal, $S$, and the noise, $N_i$, in the spatial sector. An astronomical signal estimator, $\tilde{S}_i$, was built by cleaning the raw data $m_i$, and the time-domain analogs of the $M_i$.

For extended sources, $S$ correlates with the time-stream signals as $A_i$. The atmosphere $a_i$, time-domain analogs of $A_i$ in the time sector, may not be orthogonal to the true sky signal in the time sector, $s$. An iterative technique was developed to maximize the orthogonality between the principal components being cut from time-stream data and the real astronomical signal. This means that measurement of fluxes in the initial map was subtracted from the raw data set. Then, the data streams were cleaned and mapped again. Any significant detections in the second map were added to the source model of the first generated map. The summed model was then subtracted from the initial bolometer signals. In summary, an iterative cleaning algorithm was performed using the following procedure: 1) set of $\tilde{S}_i$ was found from the set of $m_i$. $\tilde{S}$ was created by co-adding all the $\tilde{S}_i$; 2) $\tilde{r}_i$, the estimator for the time-stream data unrelated to the astronomical signal, was created by subtracting $\tilde{s}$ from $m_i$; 3) $\tilde{r}_i$ was cleaned and maps were created for the set of $\tilde{R}_i$; these were co-added to find $\tilde{R}$, the residual map of the sky signal; 4) a new estimator of the true sky signal, $\tilde{S'}$ was created by adding the residual map, $\tilde{R}$ to the map from step 1); 5) we returned to the step 2) and subtracted $\tilde{s'}$ from the input time streams. The final AzTEC image of OMC1 was the result of thirty iterations of cleaning as described above. Any induced artifacts from the iterative cleaning processes were compensated for by the inverse effects of any subtracted non-real features in the time stream data. This process was repeated thirty times until the result converged to the model-subtracted map.

\begin{figure}[h]
\plotone{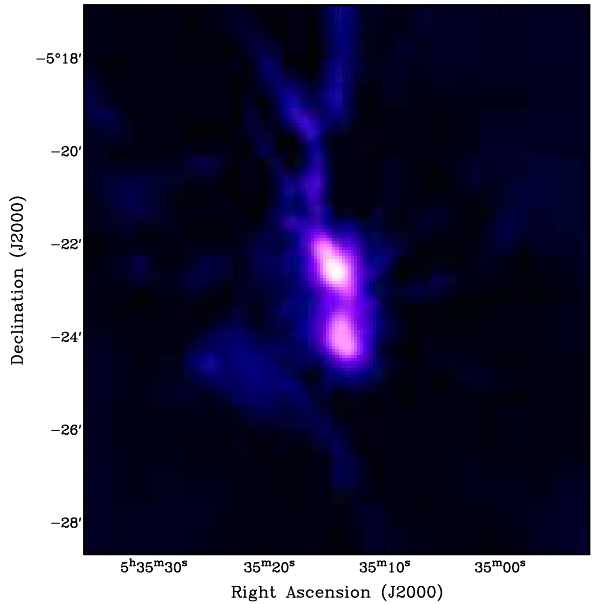}
\caption{The AzTEC 1.1mm emission from OMC1 in the Orion A molecular cloud. The peak flux of the 1.1mm emission at Orion KL of OMC1 in the Orion A molecular cloud is about 63.9 Jy/beam. The calibration error is reported as $6-13\%$ of the flux in the paper by Wilson et al. (2008). \label{fig:AzTECOMC}}
\end{figure}

\section{FOURIER ANALYSIS}

We present observations of the 1.1mm emission from OMC1 in Fig. 1. The thermal dust emission from cool dust at 1.1mm can be the result of heating from intermediate mass stars (5$-$20 M$_\odot$) and the interstellar radiation field. The morphology of the emission is similar to that of the 850 $\mu$m emission (Johnstone \& Bally 1999) seen with SCUBA at JCMT. The northern peak of OMC1 is associated with Orion KL (Kleinmann \& Low 1967; Allen et al. 1993; Menten et al. 1995; Lee \& Burton 2000; Zapata et al. 2011) in OMC1 and the southern peak of OMC1 corresponds to the Orion S source (Schmid-Burgk et al. 1990). The peak fluxes of these two sources in Fig. 1 were measured in Jy/beam and presented in Table 1.

To extract quantitative information from the statistical properties of the interstellar emission, we performed power spectrum analysis of the intensity map shown in Fig. 1. This analysis provides a useful information on understanding the physics and structure of the interstellar dust medium which was dominated by thermal emission from the cold dust in OMC1 and filaments. The parameters of the images from the power spectrum analysis were often used to study the statistical properties of pre-stellar cores in the molecular clouds and filaments as well as in the cosmic microwave background (CMB) (Shaw \& Lewis 2012) and cosmic infrared background (CIB) radiation analysis (Finkbeiner et al. 1999; Keisler et al. 2011). The power spectrum has also been used extensively in structural analysis of the images (Combes et al. 2012; see references therein). The main goal of this analysis is to place some constraints on the density structure of the interstellar dust emission arising from OMC1 using Fourier analysis (Bracewell 1986). Since dust grains are coupled with gas, elucidating the structure of the dust emission will also play an important role in understanding the dynamics and the structure of the gas. The power spectrum of the 1.1mm dust emission from OMC1 is shown in Fig. 2.

The power spectrum, $P(k)$, was measured by carrying out the Fourier transform (Bracewell 1986; Press et al. 1992) of the auto-correlation of the image (Kim et al. 1999). The Fourier transform of the product of the data with a window function reflects the convolution of the Fourier transform of the data with the window function (Bracewell 1986; Press et al. 1992). In the present analysis, a nonparametric periodogram estimator was used in the Fourier transform (Eq. 1) of the image (Fig. 1).

\begin{deluxetable}{lcc}
\tablewidth{0pt}
\tablecaption{Emission properties of Orion KL and Orion S of OMC1 in the Orion A Molecular Cloud at 1.1mm\label{table:flux}}
\tablehead{\colhead{} & \colhead{AzTEC Coordinate} & \colhead{Peak Flux} \\
 & RA[J2000]  DEC[J2000] & [Jy beam$^{-1}$] }
\startdata
 Orion KL & 05:35:14.17   -05:22:33.35 & 63.9 \\
 Orion S & 05:36:13.52   -05:24:07.35 & 26.8  \\
\enddata
\end{deluxetable}

\begin{figure}[h]
\plotone{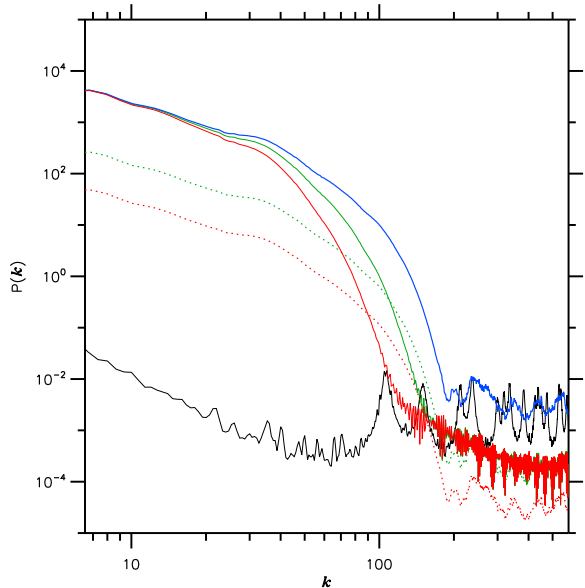}
\caption{Here, the density power spectrum of the AzTEC 1.1mm emission from OMC1 is presented. The estimate of the power spectrum was calculated with different aperture diameters of 18$^{\prime\prime}$ (blue), 36$^{\prime\prime}$ (green solid), 56$^{\prime\prime}$ (red solid), and the noise (black) for the comparison. We also compared the results of the analysis for the beam convolution with those for the averaging of the pixels (dashed lines) in the image.
\label{fig:omcpowersp}}
\end{figure}

The periodogram estimate of the power spectrum, $P(w,N)$ is defined on $N/2+1$ frequencies in Eq. 1

\begin{equation}
P(w,N)=\sum_{k=-N+1}^{N-1}{r}_{x}(k)e^{-2\pi ikw}
\end{equation}

where \begin{equation} {r}_{x}(k)=\frac{1}{N}\sum_{n=0}^{N-1-k}x(n+k)x^{\ast}(n). \end{equation}
The data were partitioned into $K$ segments and each 2$M$ consecutive points at a number of discrete frequencies. Here, $N$ is 2$M$. $k$ refers to a wavenumber. The periodogram estimates were averaged at each frequency, which reduced with the variance of the estimates by a factor of $K$. Welch's modified periodogram is justified in Eq. 3. The expected value of the periodogram converges to the power spectrum
in the mean squared sense.

\begin{eqnarray}
P(w)\approx\frac{1}{KL}\sum_{i=0}^{K-1}|\sum_{n=0}^{L-1}w(n)x(n+iD)e^{-2\pi inw}|^2
\end{eqnarray}

where the overlap occurs at $L-D$ data points and $N=L+D(K-1)$ (Press et al. 1992). $w(n)$ refers to a window.

The periodogram was calculated by computing the discrete Fourier transformation. Here, the result was an array of power measurements for each frequency bin. The amplitude of each frequency bin was determined by centering a window for an each bin and measuring the amount of signal falls within the window. The individual data sets are commonly overlapped in Welch's periodogram and this windowing of the segments makes a modified periodogram. This modified periodogram including overlap processing computes the weighted average of the sum of the spectrum from each segment with the window. The window functions usually have more influence on the data at the center of the set than on the data at the edges. In calculating the periodogram, 50\% of overlapping is treated in this paper.

The variance of the periodogram estimate can be defined as
\begin{eqnarray}
\sigma^2=\frac{1}{L}\sum_{n=0}^{L-1}|w(n)-\mu|^{2}
\end{eqnarray}
where $\mu$ refers to a mean. When the segments are overlapped by one half of their lengths, the variance of Welch's periodogram reaches an optimal value. This reduces the variance by a factor of about 0.818$K$ (Press et al. 1992). If there is an increase in the number of subsequences to be averaged, then the variance of the estimate is reduced significantly. Therefore, we can say that the power spectrum estimation is approximately unbiased.

\section{POWER SPECTRA}

The power spectrum of density fluctuations can be related to the power spectrum of intensity fluctuations (Bondi et al. 1994). In this analysis, we found that it was difficult to fit a single power law to the observed spectrum in Fig. 2. The power law at large scales is shallow because it connects to the large-scale portion, while the power spectrum is dominated by the size and shape of the map. The power spectrum obtained from the intensity map of OMC1 in the present analysis displays at least three distinct power law exponents as can be seen in Fig. 2. The best-fit slope in the range $30 \lesssim k \lesssim 100$ is $\gamma\approx-2.66\pm0.3$. This result is comparable to the spectral index $\gamma\approx-2.7$ that was found in numerical simulations (Padoan et al. 2004) within error estimates. There are distinct spectral breaks at spatial frequencies of $k\approx30$ and at $k\approx100$ possibly. While the break in the power spectrum occurring at the thickness of the galactic disk was noted in Elmegreen, Kim, \& Staveley-Smith (2001), the preferred scale apparent in the power spectrum was also discussed by Brunt (2010) and in the previous studies by Blitz and Williams (1997) and Hartmann (2002).

At smaller scales the observed power spectrum of the image representing the 1.1mm emission from OMC1 approaches the noise level.
The power spectrum steepens at relatively smaller scales. The power spectrum approaches the noise level at higher spatial frequency (Fig. 2).
Estimates of the power spectra at smaller scales can often underestimate the value, since structures can be suppressed below the sonic scale (V\'azquez-Semadeni, Ballesteros-Paredes, \& Klessen 2003). From the present analysis (Fig. 2), the power spectrum steepens at relatively smaller scales. The logarithmic slopes of the power-law spectra obtained at low spatial frequency seem to be shallower than the logarithmic slope of Kolmogorov spectrum. A Kolmogorov spectrum has a power law with a logarithmic slope of $-11/3$, conventionally taken as evidence for turbulent fluctuations in dynamical energy exchange among irregularities of different scales.
The spectral indices of the power spectra of the dust emission derived from other studies (Gautier et al. 1992) are also rather shallower than that of Kolmogorov spectrum. Most spectral indices measured in other studies such as Cirrus cloud studies and Cassiopeia A (Roy et al. 2008) range from $-3.6$ to $-2.2$. But in the present Fourier analysis, the slope of the power law at high spatial frequency becomes steeper than the logarithmic slope of Kolmogorov spectrum. Often spectra with logarithmic slope of four or more are considered to be caused by deterministic structures. However, it is difficult to explain the power source for forming these substructures.

Recent studies of turbulence indicate a rather shallow slope of the density power spectrum. For example, Kritsuk \& Norman (2004), Beresnyak, Lazarian, \& Cho (2005), Kim \& Ryu (2005), and Bournaud et al. (2010) found slopes which were rather smaller than 2 from their magnetohydrodynamic (MHD) turbulence. The slopes of power law become smaller as the Mach number increases in their simulations and measurements. In the present Fourier analysis, the slope of the power law at low spatial frequency, $k\lesssim30$ is $\gamma\approx-1.5\pm0.1$ and smaller than 2. In the transonic turbulence model, the density distribution is often generated by discontinuities created by weak shocks overlaid on the turbulent background. In the supersonic turbulence, the density distribution is often characterized as peaks, or mass concentrations generated by strong shocks (Williams, Blitz, \& McKee 2000). As the Mach number increases, the power spectrum flattens (Kim \& Ryu 2005), and density concentrations appear as sheets and filaments (McKee \& Ostriker 2007). To interpret these results in terms of the density structure (Brunt \& Heyer 2002) and in comparison with deductions from gas tracers, we reproduced the results by considering a model spectrum based on the 1.1mm emission which is dominated by the gray-body emission from large dust grains at thermal equilibrium temperature (D$\acute{e}$sert et al. 1990). Here, the large grain equilibrium temperature is related to the local radiation field strength (Stepnik et al. 2003) and spectrum which depends on the presence or absence of nearby heat sources as well as extinction. Since grain structure variation and size can affect the emissivity coefficient and can result in a variation of the equilibrium dust temperature, this will consequently constrain the pattern of the power spectra at smaller and intermediate scales in the interstellar dust medium.

To understand the observed shape of the power spectrum, we attempted to analyze the impact of variations in the beam size on the pattern of the observed power spectrum of the OMC1 image. We performed power spectrum calculations using three different aperture diameters: 18$^{\prime\prime}$, 36$^{\prime\prime}$, and 56$^{\prime\prime}$. Here, the aperture diameter is an average for the observed region. The pixels in the data have been averaged using {\rm IMBIN} task in {\rm MIRIAD} (Sault et al. 1995). Each resolution of the newly derived map, 36$^{\prime\prime}$ and 56$^{\prime\prime}$ was achieved by averaging 9 and 14 pixel squares respectively. Each new pixel was replaced with the average value of the pixels within the resolution of the map. Despite differences in the details, the power spectra of the 1.1mm emission at apertures of different sizes are similar in shape. However, as the size of aperture increases, the power decreases significantly (Fig. 2). The power spectrum is slightly less steep towards the large beam at relatively small scales, compared to the large scale structure of OMC1 at 1.1mm. The slope of the observed power spectrum is more affected after beam convolution than after averaging over the pixel values. For beam convolutions, a gaussian beam was used to convolve the image at different resolutions using the {\rm CONVOL} task in {\rm MIRIAD}. The power spectra at lower and higher resolutions calculated from the beam convolution have similar appearances at spatial frequencies smaller than $k\approx30$. But the slope changes significantly at higher spatial frequency as the aperture diameter increases. The average error in the noise was also measured in terms of the spatial power spectrum, and is presented in Fig. 2. The noise includes 1/$f$ noise (Wilson et al. 2008) and confusion noise from the background. The shape of the power spectrum of the noise follows a power law and the present analysis yields a spectral index of $\gamma \approx-0.6$.

\section{SUMMARY}

We present a 1.1mm emission map of the OMC1 region observed with AzTEC, a new large-format array composed of 144 silicon-nitride micromesh bolometers that was in use at the JCMT. We performed a Fourier analysis of the image with a modified periodogram. The periodogram was measured by computing the discrete Fourier transformation. From the present analysis of the OMC1 filaments at 1.1mm emission, the power spectrum steepens at relatively smaller scales. At larger scales, the power spectrum flattens and the large scale power law becomes shallower. The logarithmic slopes of the power-law spectra obtained seem to be lower than the logarithmic slope of Kolmogorov spectrum. We also performed power spectrum calculations using three different aperture diameters. As the size of aperture increases, the power decreases significantly. We were able to fit at least three components of power law in the power spectrum of the 1.1mm emission map. The slope of the best fit at spatial frequency of $30 \lesssim k \lesssim 100$ is $\gamma\approx-2.66\pm0.3$. This result is similar to the spectral index of the power spectrum, $\gamma\approx-2.7$ that was found in numerical simulations. The effects of noise on the slope of the power spectrum were also included in the present analysis.

\acknowledgments
{The James Clerk Maxwell Telescope is operated by the Joint Astronomy Center on behalf of the Science and Technology Facilities Council of the United Kingdom, the Netherlands Organization for Scientific Research, and the National Research Council of Canada. We appreciate for helpful comments and help from the AzTEC team and other colleagues. We thank an anonymous referee for helpful comments. This work was supported in part by the faculty research fund of Sejong University. SY was supported in part by Mid-career Researcher Program through the National Research Foundation of Korea (NRF) funded by the Ministry of Education, Science, and Technology 2011-0028001.}


\end{document}